\documentclass[jctc,manuscript=article]{achemso}

\makeatletter

\def\@seccntformat#1{\csname the#1\endcsname\quad}
\makeatother

\usepackage{graphicx}
\usepackage{amsmath,amssymb,hyperref}
\usepackage{xcolor}
\usepackage{booktabs}
\usepackage{siunitx}
\usepackage{braket}
\DeclareSIUnit\angstrom{\text{Å}}
\title{Quantum-Centric Geometry Optimization with Wave-Function-Based Embedding}

\author{Danil Kaliakin}
\affiliation{Computational Life Sciences, Cleveland Clinic Research, Cleveland, Ohio 44106, United States}

\author{Akhil Shajan}
\affiliation{Computational Life Sciences, Cleveland Clinic Research, Cleveland, Ohio 44106, United States}

\author{Fangchun Liang}
\affiliation{Computational Life Sciences, Cleveland Clinic Research, Cleveland, Ohio 44106, United States}

\author{Zhen Li}
\affiliation{Computational Life Sciences, Cleveland Clinic Research, Cleveland, Ohio 44106, United States}

\author{Kenneth M. Merz Jr.}
\affiliation{Computational Life Sciences, Cleveland Clinic Research, Cleveland, Ohio 44106, United States}
\alsoaffiliation{Department of Chemistry, Michigan State University, East Lansing, Michigan 48824, United States}
\email{kmerz1@gmail.com}

\abbreviations{EWF,SQD,PES}

\begin{document}

%=================================================================
\begin{abstract}
The EWF-(FCI,SQD) method — a wave-function-based embedding approach combining full configuration interaction (FCI) and sample-based quantum diagonalization (SQD) — is a promising new tool for the simulation of molecular systems. However, applications of EWF-(FCI,SQD) have so far been limited to single-point calculations, whereas the study of complex chemical processes requires the ability to explore potential energy surfaces. In this work, we demonstrate geometry optimization with EWF-(FCI,SQD), scaling our simulations to molecules as large as menthone and benzidine within the STO-3G basis set. Without fragmentation, these systems comprise 73 and 82 molecular orbitals respectively, presenting an intractable Hilbert space for conventional exact or high-level subspace solvers and establishing a clear necessity for fragmentation-based methodologies. The underlying fragment SQD simulations in the EWF-(FCI,SQD) geometry optimizations use up to 70 qubits. The resulting geometries show exceptional accuracy relative to the classical reference, with deviations below 4 picometers.
\end{abstract}

%=================================================================
\section{Introduction}

Quantum-centric computing, in which classical high-performance computing (HPC) and quantum processing units (QPUs) operate synergistically,\cite{seelam2026reference} is a rapidly growing paradigm in the field of electronic structure simulation. Quantum-centric computation of the electronic structure of molecular systems is currently realized through the quantum-selected configuration interaction (QSCI) method \cite{qsci_ref} and its variant, the sample-based quantum diagonalization (SQD) \cite{robledo2024chemistry,shirakawa2025closedloopcalculationselectronicstructure}.

While the majority of quantum-centric studies of electronic structure have been focused on single-point energy calculations,~\cite{sqd_original,sqd_inter_mol,Ieva2025,nutzel2024solving,kaliakin2025implicit,barison2024ext-sqd,shirakawa2025closedloopcalculationselectronicstructure,sugisaki2025hamiltonian,piccinelli2025quantum,shirai2025enhancing,sugisaki2025size,Shirai2025,barroca2025surface,ieva_diazo,nogaki2025symmetry,wray2025convergence,raisuddin2025promise} practical computational chemistry simulations exploring the potential energy surfaces, such as molecular dynamics and geometry optimization,\cite{Badar2022,global_opt,shajan2023geometry} require evaluating the electronic energy gradient with respect to the nuclear positions of the molecules under study. SQD studies show promise in electronic gradient simulations, with early successes demonstrated in the molecular dynamics of small molecules and the active sites of proteins.\cite{bazayeva2025quantum,li2026protein}

Scaling SQD simulations beyond small molecules and protein active sites can be achieved with fragment-based methods.\cite{wang2026localized,shajan2025toward,trp_cage,sqd_bootstrap_dmet,patra2025quantum} Wave-function-based embedding (EWF)\cite{Booth2023} combined with SQD\cite{trp_cage,das2026quantumcomputationsfusionblanket} has proven particularly promising in this regard: EWF-(FCI,SQD) simulations, where a full configuration interaction (FCI) solver treated simulations of hydrogen atom fragments and a SQD solver was utilized for atoms heavier than hydrogen, have been applied to a protein system of over 12,000 atoms.\cite{merz2026crossing12000atombarrierheterogeneous} Nonetheless, despite substantial progress in the scalability of EWF-SQD simulations, this method has so far been applied only to single-point energy calculations, while electronic gradient simulations remain limited to conventional, unfragmented SQD.\cite{bazayeva2025quantum,li2026protein}

In this work, we demonstrate that the EWF-SQD method can be efficiently applied to the geometry optimization of systems from the Baker test set\cite{Baker_test_set} of organic molecules. Although density matrix embedding theory (DMET)—of which EWF is an extension—has been used in quantum computing simulations before, this study was carried out within the variational quantum eigensolver (VQE) paradigm.\cite{hao2026large} To our knowledge, no geometry optimization with EWF-SQD has been reported to date. Whereas the largest molecule treated in the previous DMET-VQE study contained 9 atoms,\cite{hao2026large} here we extend fragment-based geometry optimization with quantum computing algorithms to a molecular system of as many as 29 atoms. Similarly to previous DMET-VQE study\cite{hao2026large} we perform our simulations with the sto-3g basis set. This minimal basis set is an appropriate choice for an initial demonstration of the feasibility of the new methodology, while its scalability to more practical basis sets will be addressed in future studies.

To establish the accuracy of EWF simulations, we first evaluate the method with a selected configuration interaction (SCI) solver,\cite{SCIEriksen,modern_sci} benchmarking {EWF-(FCI,SCI)} results against unfragmented SCI calculations.
In all EWF calculations reported in this work, fragments with active spaces of 12 or fewer molecular orbitals are solved exactly with FCI rather than with the nominal fragment solver; we denote this mixed-solver protocol EWF-(FCI,SCI) or EWF-(FCI,SQD), following the notation of our previous EWF studies, according to the approximate solver assigned to larger fragments. This comparison uses the first ten molecules from the Baker test set: water, ammonia, ethane, acetylene, allene, hydroxysulfane, benzene, methylamine, ethanol, and acetone.\cite{Baker_test_set} Having isolated the effect of EWF fragmentation itself, we then compare EWF-(FCI,SCI) and EWF-(FCI,SQD) for the geometry optimization of these molecules. To probe the scalability of EWF-(FCI,SQD), we apply it to menthone and benzidine, the two molecules with the largest number of atoms in the Baker test set,\cite{Baker_test_set} and confirm its accuracy by comparing the optimized geometries with those obtained from EWF-(FCI,SCI).

In addition to extending the fragment-based quantum computing geometry optimization methodology from DMET-VQE to EWF-SQD and targeting substantially larger molecules, we introduce several other important innovations, discussed below. First, whereas the previous DMET-VQE study\cite{hao2026large} was demonstrated on classical simulators of quantum circuits, the present work performs geometry optimization with real quantum hardware. Second, the previous numerical experiments showed that DMET-VQE could be scaled to a largest DMET fragment of 20 qubits, corresponding to 10 molecular orbitals (MOs);\cite{hao2026large} here we show that EWF-SQD can be scaled to a largest fragment of 33 MOs, which in the SQD method corresponds to 70 qubits. Third, to our knowledge, this is the first study in which a GPU-accelerated SBD eigensolver is integrated into the reference classical SCI calculations, which previously was only integrated within the framework of truncated CI.\cite{das2026quantumcomputationsfusionblanket} This is an important step toward more efficient reference calculations that enabled us to perform unfragmented SCI on molecules as large as benzene. Finally, we incorporate the EWF density response into the nuclear gradient by solving the Lambda equation of cluster amplitudes,\cite{cc_in_quant_chem} which further improves the stability of the gradient calculations.

The Methods section shows the interaction between the software components used to build the EWF geometry optimization workflow and describes: the EWF embedding methodology and fragment construction, the configuration interaction (CI) solvers employed in this work, the generation of candidate electronic configurations from quantum hardware together with the associated classical post-processing in SQD, the incorporation of the EWF density response into the nuclear gradient via the $\Lambda$ equations,\cite{cc_in_quant_chem} the resulting analytical EWF gradient, and the Sella geometry optimizer\cite{sella_opt}. We choose the Sella geometry optimizer as the sole optimizer used in this work, because previous study demonstrated it as a most efficient choice among the publicly available open source optimizers.\cite{shajan2023geometry} The Results section first evaluates the performance of EWF-(FCI,SCI) geometry optimizations against unfragmented SCI references. The agreement between the two methods is quantified using the root-mean-square and maximum deviations of the optimized geometries. We then demonstrate how closely EWF-(FCI,SQD) reproduces the EWF-(FCI,SCI) geometry optimization results. For all EWF simulations, we report the total number of atoms in each simulated system (which also equals the number of fragments) and the number of MOs in the largest EWF fragment of that system. For the EWF-(FCI,SQD) simulations, we additionally report the number of qubits used for the largest fragment in each system. The Conclusions section highlights key advances, present limitations, as well as future directions for quantum-centric geometry optimization and molecular dynamics with EWF.

%=================================================================
\section{Methods}

\begin{figure}[ht]
\centering
\includegraphics[width=\linewidth]{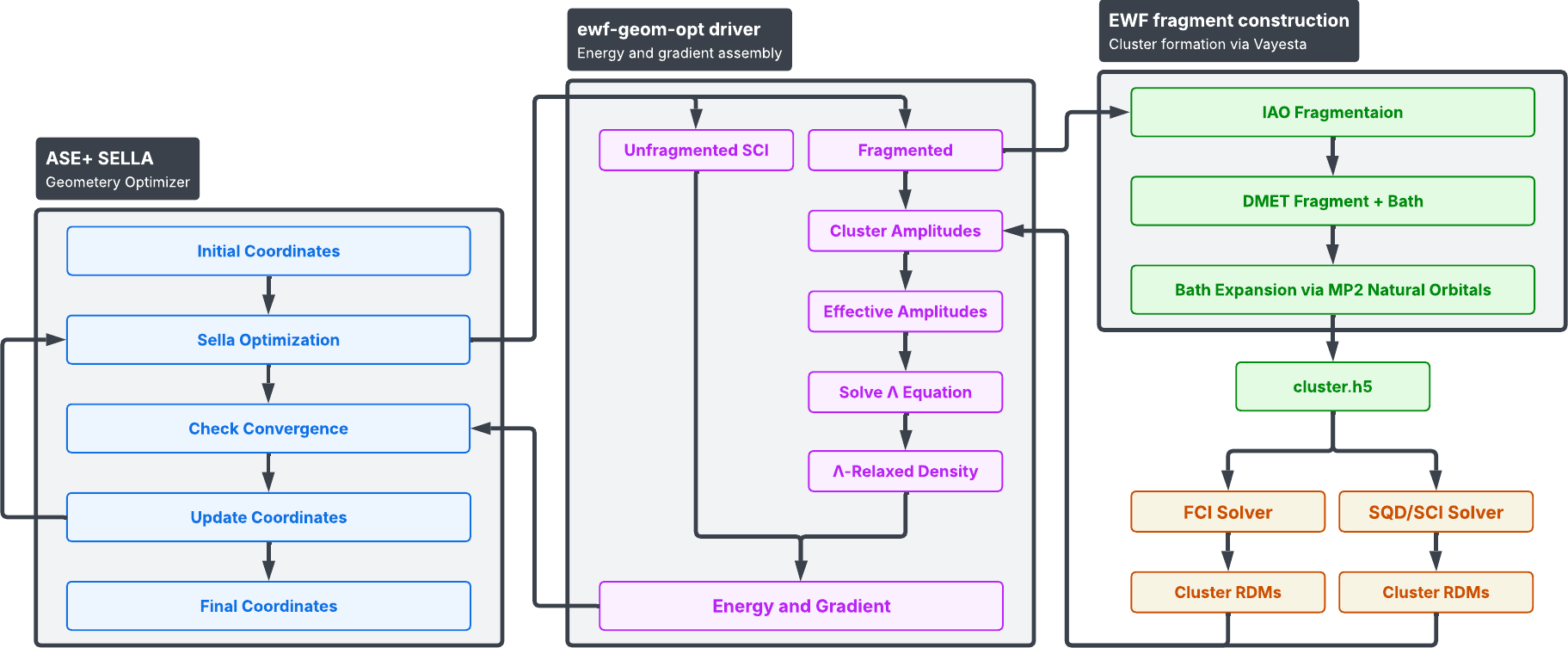}
\caption{\textbf{Overall workflow of the EWF geometry optimization framework.} ASE together with Sella drives the geometry optimization by repeatedly requesting the electronic energy and analytical gradient from the geometry driver. For fragmented calculations, EWF clusters are constructed in the IAO basis using Vayesta, expanded with MP2 natural orbitals, and solved using FCI or SQD/SCI depending on the fragment size. The resulting cluster reduced density matrices are converted into cluster and effective amplitudes, from which the $\Lambda$ equations are solved to construct the $\Lambda$-relaxed density; the analytical gradient is then evaluated from this density alongside the assembled EWF energy, which is returned to the optimizer. Unfragmented SCI calculations follow the same optimization workflow, without fragmentation or $\Lambda$-density relaxation, providing the reference calculations used throughout this work.}
\label{fig:workflow}
\end{figure}

Figure~\ref{fig:workflow} summarizes the complete geometry optimization
workflow developed in this work. Geometry optimization is performed using
ASE together with Sella, while the EWF geometry driver evaluates the
electronic energy and analytical gradient through either fragmented or
unfragmented electronic structure calculations. The individual components
of this workflow are described in the following subsections. The associated code is distributed as one of the repositories in MIQuLab  and is fully open-source~\cite{ewf-geom-opt-repo}.

\subsection{Wave Function-Based Embedding (EWF)}
Within the workflow shown in Figure~\ref{fig:workflow}, EWF fragments the system into atomic clusters defined in the intrinsic atomic orbital (IAO) basis (Figure~\ref{fig:workflow}). Each fragment comprises an atom together with a set of bath orbitals, obtained from a Schmidt decomposition of the mean-field one-particle density matrix, that capture the entanglement between the atom and its environment.\cite{nusspickel2022systematic} The bath is expanded with local MP2 natural orbitals, retaining virtual (occupied) orbitals with occupation numbers above $\eta$ (below $2-\eta$): decreasing $\eta$ enlarges each fragment, and as $\eta\to0$ the EWF description converges to the full-molecule result.\cite{nusspickel2022systematic,trp_cage} We use $\eta = 1\times10^{-5}$ for all fragments in this work. Each resulting EWF cluster is solved at a correlated level with FCI, SCI, or SQD; independently of $\eta$, the solver assigned to a given fragment follows a separate size-based rule, with clusters of 12 or fewer active MOs solved exactly by FCI, consistent with the mixed-solver protocol stated in the Introduction. The correlated one- and two-particle density matrices from each cluster are projected back into the global MO basis and combined across fragments into the assembled one- and two-particle densities $\gamma_1$ and $\lambda_2$, from which our workflow evaluates the total EWF energy as the sum of the Hartree--Fock energy, the one-particle correlation correction contracted with the Fock matrix, and the two-particle cumulant contracted with the electron-repulsion integrals.

\subsection{Configuration Interaction Solvers}
\subsubsection{GPU-enabled Selected Basis Diagonalization}

We use the selected basis diagonalization (SBD), developed by RIKEN\cite{shirakawa2025closedloopcalculationselectronicstructure} and recently enabled on GPUs \cite{doi2026gpuacceleratedselectedbasisdiagonalization}, as the underlying GPU-accelerated eigensolver for SCI and SQD simulations in the present study. First, the quantum-hardware- or SCI-generated bitstrings are generated to represent the set of selected configurations and serve as the input data for the SBD eigensolver. Second, the selected configurations $\mathcal{S}$ define the distributed wavefunction vector, which is described in eq.\eqref{wavefunction}. Third, to solve the eigenvalue and obtain the energy of the selected configurations, a GPU-based Davidson diagonalizer is introduced. 

\begin{equation}
\boldsymbol{\psi}_{\mathcal{S}}
=
\left(
\psi_{\mathbf{x}_1},
\psi_{\mathbf{x}_2},
\ldots,
\psi_{\mathbf{x}_{|\mathcal{S}|}}
\right)^{\mathrm T}.
\label{wavefunction}
\end{equation}

To establish the parallel reading of bitstrings from the input file stored on disc, the MPICH 5.0.1 \cite{mpich501} library compiled by llvm 18.1.8 clang++ compiler \cite{clang1818} is used for message broadcasting and reducing between multiple CPUs. After the bitstring has been saved in cache, the subsequent communication between CPU-read bitstrings and GPU processors needs CUDA toolkit 12.3 \cite{cuda123}. Finally, the Davidson diagonalization performed on each GPU processor is supported by the OpenBlas linear algebra package \cite{openblas0330dev}. All compilation and testing of GPU-SBD in this work was performed on the RHEL system of Cleveland Clinic's High Performance Computing Center using 4-blade NVIDIA A100 GPU nodes with 512 GB RAM and 40 GB VRAM. To orchestrate individual executions of the SBD eigensolver we utilize the SLURM job scheduling software\cite{slurm} controlling execution of each SBD job.

\subsubsection{Selected Configuration Interaction}

To reuse the highly optimized determinant-based FCI framework (specifically, the alpha/beta string decoupling), the SCI solver in PySCF\cite{sun2018pyscf,sun2020recent} (\texttt{pyscf.fci.selected\_ci}) restricts its variational subspace $\mathcal{V}_{\text{SCI}}$ to be the strict Cartesian product of a selected subset of $\alpha$-strings $\{\alpha_i\}$ and $\beta$-strings $\{\beta_j\}$:
\begin{align}
    \mathcal{V}_{\text{SCI}} = \left\{ |\alpha_i\rangle \otimes |\beta_j\rangle \right\}.
\end{align}
Consequently, if a specific configuration $|\alpha_i\beta_j\rangle$ is variationally selected, all unselected combinations within the cross product (e.g., $|\alpha_i\beta_k\rangle$) are artificially forced into the active space. This architectural constraint introduces an inefficient growth of trivial determinants with near-zero amplitudes, generating a highly structured but sparse Hamiltonian matrix over the tensor product space.
In the Selected Configuration Interaction (SCI) stage, the truncation of the variational subspace is governed by the selection threshold inside the PySCF \texttt{selected\_ci} kernel. Specifically, we employ the standard PySCF default threshold, defined via the configuration parameter \texttt{sci\_select\_cutoff}: $\varepsilon_{\text{SCI}} = 1.0 \times 10^{-3}.$
This cutoff serves as the strict entry criterion during the subspace expansion, controlling the perturbative filtering of external configurations. 
Only those single and double excitations that exhibit a first-order energetic contribution greater than $\varepsilon_{\text{SCI}}$ are retained.
Once the final variational subspace is established by the selection kernel, the standard CPU-based diagonalizer in PySCF is bypassed. Instead, the final subspace Hamiltonian matrix is passed to our GPU-accelerated SBD solver. By offloading the large-scale sparse matrix-vector multiplications to GPU architectures, the SBD solver achieves significantly faster eigenvalue resolution while maintaining exact variational equivalence to the native PySCF implementation.

\subsubsection{Sample-Based Quantum Diagonalization}
In EWF-(FCI,SQD) we employ the sample-based quantum diagonalization (SQD)\cite{robledo2024chemistry} approach to treat the fragments with larger active spaces (more than 12 MOs). While for systems with small number of MOs the FCI calculations are trivial and inexpensive on modern CPUs, at the larger MO counts the exact FCI becomes dramatically more computationally expensive and eventually prohibitive.\cite{Vogiatzis2017,modern_sci,Fales2020,Gao2024} 
SQD operates as a hybrid quantum-classical workflow designed to systematically truncate the active-space Hilbert space for classical diagonalization while maintaining high chemical accuracy.\cite{robledo2024chemistry}

The workflow begins by sampling electronic configurations from the local unitary cluster Jastrow (LUCJ) circuits\cite{motta2023bridging} implemented on a quantum processor. 
To mitigate device noise that yields configurations with an incorrect number of particles, we apply a classical post-processing procedure termed configuration recovery to each sample batch\cite{robledo2024chemistry}. This procedure identifies and transforms non-physical configurations, redistributing them into physically valid states based on their statistical behavior.
To further refine the energy and property predictions, an extended SQD (ext-SQD) scheme is invoked\cite{barison2024ext-sqd}. Ext-SQD selectively extracts dominant configurations from the final, lowest-energy batch. 
This core subspace is then expanded by applying all single-excitation operators, followed by a final classical diagonalization. We perform the application of single-excitation operators using the software stack enabled through the PyCI software package.\cite{richer2024pyci}

\textit{\textbf{Local Unitary Cluster Jastrow (LUCJ) Ansatz and Circuit Execution:}}
The quantum circuits utilized for sampling candidate configurations within the SQD framework are parameterized via the Local Unitary Cluster Jastrow (LUCJ) ansatz \cite{motta2023bridging}. 
Under the Jordan-Wigner (JW) transformation, $M$ spatial orbitals are mapped onto $2M$ qubits. 
To suppress noise accumulation, the parameterized LUCJ circuit is restricted to a single layer ($L=1$) in this work.
Starting from a closed-shell Hartree-Fock reference state ($\Psi_0$) initialized by a layer of Pauli-$X$ gates, the target state is prepared via the following state evolution:
\begin{align}
    \ket{\Psi_{LUCJ}}=\prod_{\mu=0}^{L-1} e^{\hat{K}_\mu} e^{i\hat{J}_\mu} e^{-\hat{K}_\mu} \ket{\Psi_0},
\end{align}
where the unitary operators are constructed to comply with hardware topology constraints:
\begin{itemize}
    \item Orbital Rotations ($e^{\pm\hat{K}_\mu}$): Implemented via a dense brickwork pattern of Givens rotation gates optimized for 1D qubit connectivity, followed by a single layer of single-qubit phase gates to finalize the basis transformation.
    \item Density-Density Interactions ($e^{i\hat{J}_\mu}$): Encoded as a combination of single-qubit phase gates and a sequence of two-qubit Controlled-Phase gates, which parameterize the Jastrow-type electronic correlations.
\end{itemize}
To render the circuit compatible with NISQ-era hardware limitations, a locality approximation is introduced. Interactions between spin-orbitals that do not map to geometrically adjacent qubits under the JW transformation are strictly truncated to zero. 
This localization enables a SWAP-gate-free implementation, restricting the total gate count to scale linearly with the number of qubits while maintaining a constant circuit depth for the interaction blocks.

The initialization and optimization of the circuit parameters, as well as the circuit construction, are entirely managed using the \texttt{ffsim} software package\cite{sung2026ffsim}.
The initial parameters are obtained by performing a compressed double factorization of the classical Coupled Cluster Singles and Doubles (CCSD) $t_2$ amplitudes:
\begin{align}
    \overline{t}_{ij}^{ab} = i\sum_{\mu=0}^{L-1}\sum_{pq} J_{pq}^{(\mu)} U_{ap}^{(\mu)} U_{ip}^{(\mu)*} U_{bq}^{(\mu)} U_{jq}^{(\mu)*},
\label{eq:tbar}
\end{align}
where each $U^{(\mu)}$ is an orbital rotation and each $J^{(\mu)}$ is a Coulomb matrix. 

Prior to hardware execution, these gate parameters are optimized via a single-stage classical optimization using the L-BFGS-B algorithm to minimize the residual norm between the truncated double-factorized $t_2$ amplitudes and the original CCSD target:\cite{Lin_2025_ImprovedParameterInitialization}
\begin{align}
    \chi = \frac{1}{2}\sum_{ijab}|\bar{t}_{ij}^{ab}-t_{ij}^{ab}|^2.
\end{align}

\begin{figure}
    \centering
    \includegraphics[width=0.5\linewidth]{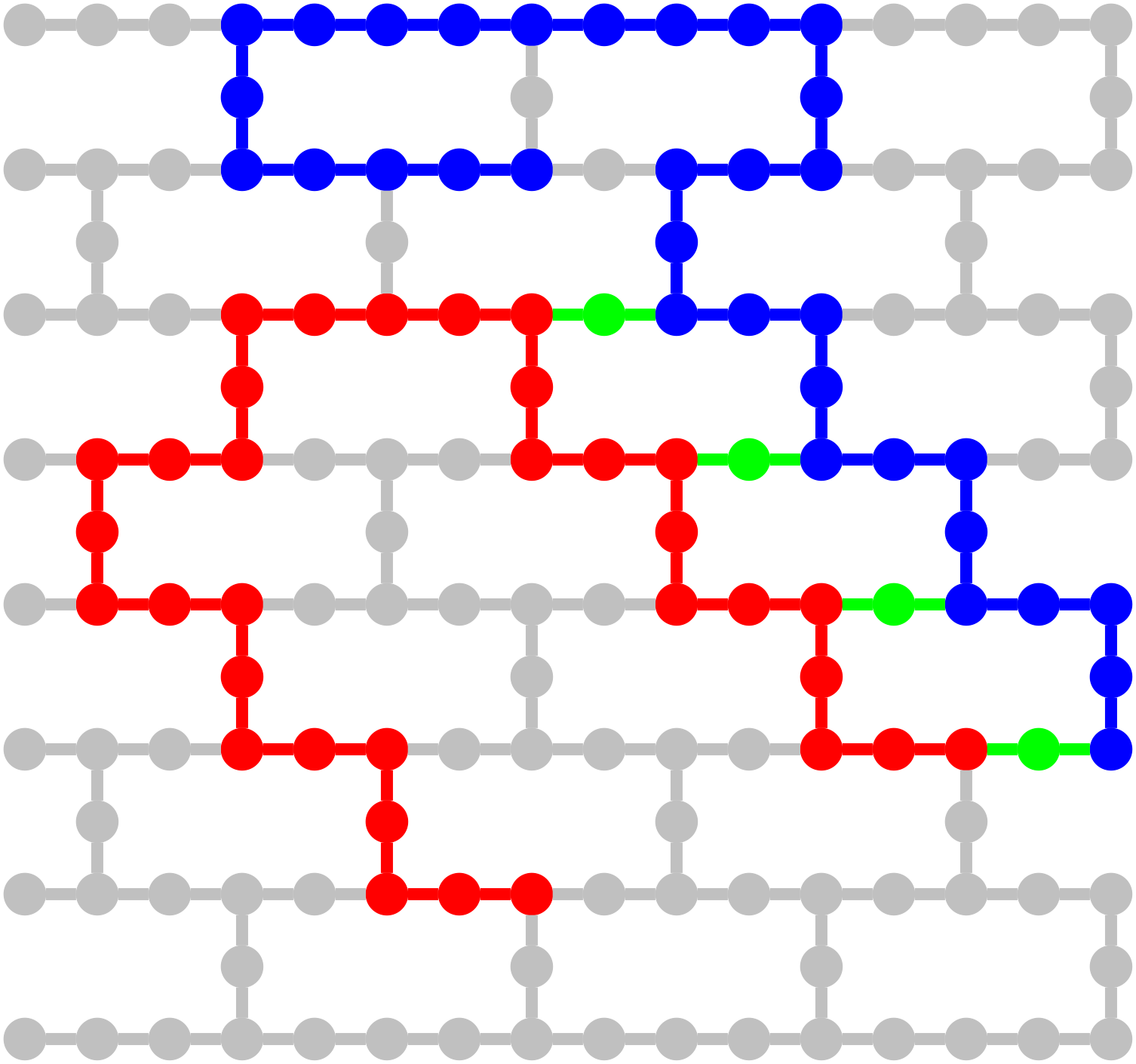}
    \caption{Example qubit layout for the largest EWF cluster in benzidine (33 molecular orbitals). Red and blue dots denote $\alpha$- and $\beta$-qubits, respectively, while green dots represent the four ancillary qubits connecting the $\alpha$–$\beta$ pairs.}
    \label{fig:qubit_layout}
\end{figure}
The optimized circuits are then constructed and executed on the IBM Quantum Processing Units (QPUs) \texttt{ibm\_cleveland} and \texttt{ibm\_marrakesh}, both of which are 156-qubit IBM Heron-r2 processors. The \texttt{ibm\_marrakesh} is utilized in simulations of benzidine and menthone, while \texttt{ibm\_cleveland} is used for simulations of all other systems in the present study. To suppress gate and coherence errors during hardware execution, a dynamical decoupling protocol employing the "XY4" sequence\cite{ezzell2023dynamical} is applied. 
As shown in Figure \ref{fig:qubit_layout}, we adopted a zigzag layout for the 33-orbital EWF cluster, utilizing a maximum of four ancillary qubits to connect the $\alpha$–$\beta$ pairs. To minimize hardware overhead and decoherence, the exact device layout was selected using a heuristic function based on current qubit error profiles reported on the IBM Quantum Platform. For each LUCJ circuit we collect 100,000 measurements (shots).

Following hardware execution, the collected bitstring counts are fed directly into the classical post-processing pipeline. 
To enhance sample quality, the carryover procedure implemented in \texttt{qiskit-addon-sqd}\cite{qiskit_addon_sqd} is utilized with a weight threshold $1\cdot10^{-4}$, preserving the most dominant configurations across successive recovery steps. 
Within the SQD workflow, the maximum number of configuration recovery iterations is set to 5, with the convergence thresholds for the total energy and average orbital occupancy established at $1\cdot10^{-8}$ and $1\cdot10^{-5}$, respectively.

Upon convergence of the core configuration recovery iterations, the ext-SQD\cite{barison2024ext-sqd,barroca2025surface} scheme is applied to broaden the subspace. Specifically, dominant configurations with a CI coefficient threshold of $c > 1\cdot10^{-5}$) are extracted from the final, lowest-energy sample batch and systematically expanded by applying all single-excitation operators. Finally, the projected Hamiltonian is classically diagonalized within this expanded, yet drastically reduced,\cite{barroca2025surface} subspace to yield the refined electronic ground-state energy.
The resulting wavefunction is further employed to compute the reduced density matrices (RDMs) required for the subsequent embedding environment calculations.

\subsection{Analytical Gradient}

\subsubsection{Gradient Formulation}
The nuclear gradient of the EWF energy follows the usual chain-rule split between the explicit and implicit dependence of the assembled density $\gamma=(\gamma_1,\lambda_2)$ on the nuclear geometry $x$,
\begin{equation}
\frac{dE}{dx} = \left.\frac{\partial E}{\partial x}\right|_{\gamma\ \mathrm{fixed}} + \left\langle \frac{\partial E}{\partial \gamma},\ \frac{d\gamma}{dx}\right\rangle
\label{eq:split}
\end{equation}
The first term is the frozen-density gradient: the derivative of the integrals at fixed $\gamma_1$, $\lambda_2$, including the Hartree--Fock coupled-perturbed (CPHF) orbital response.\cite{pople1979derivative} The second term is the density-response contribution, arising because $\gamma$ is itself rebuilt from the fragment solutions at every geometry. For a variational wave function this term vanishes identically, since the density extremizes the energy for the given integrals (Hellmann--Feynman theorem)\cite{feynman,politzer2018hellmann}. The EWF density does not have this property, because it is assembled by projecting independent fragment solutions rather than by extremizing the global energy functional directly, the same structural situation encountered in MP2 and coupled-cluster theory, where the wave function is likewise non-variational, so the density-response term is generally nonzero and must be evaluated explicitly.

\subsubsection{Density Response}
Evaluating the amplitude-response term of Eq.~\ref{eq:split} directly would require differentiating each fragment's cluster amplitudes with respect to the nuclear coordinates, in principle re-solving every fragment for each of the $3N$ displaced geometries. We avoid this with a Lagrangian (Z-vector) construction: the amplitude equations are instead augmented with adjoint multipliers, chosen so that the resulting Lagrangian is stationary with respect to the amplitudes, reducing the response evaluation to a single additional linear solve on the assembled global wave function, independent of $3N$.

\subsubsection{The $\Lambda$ Equations}
For each fragment, the projected correlated one- and two-particle density matrices are reinterpreted as an effective set of singles and doubles amplitudes on the assembled Hartree--Fock reference, $T_1^{\mathrm{eff}} = (\Delta\gamma_1)_{\mathrm{ov}}$ and $T_2^{\mathrm{eff}} = (\lambda_2)_{\mathrm{oovv}}$, projected onto the fragment's occupied index and rotated into the global molecular-orbital basis before being accumulated across fragments into a single global effective amplitude pair $(T_1,T_2)$. Together with the assembled Hartree--Fock reference, this amplitude pair defines an effective global wave function. A common simplification is to set the adjoint multipliers equal to the amplitudes themselves, $\Lambda = (T_1,T_2)$; this linearized choice does not solve the adjoint ($\Lambda$) equations and therefore captures the amplitude response only approximately.\cite{cc_in_quant_chem} 

In the present study we instead solve the $\Lambda$ equations for the fixed amplitude pair to obtain the true adjoint multipliers. The resulting relaxed one- and two-particle densities, $\gamma_1[T,\Lambda]$ and $\lambda_2[T,\Lambda]$, are then passed to the same integral-derivative contraction used to evaluate the frozen-density term of Eq.~\ref{eq:split}. Substituting the relaxed densities in place of the unrelaxed ones is what reproduces the amplitude-response contribution to $dE/dx$, following the generalized Hellmann--Feynman construction of coupled-cluster analytic gradient theory, with no separate response term evaluated explicitly.

\subsubsection{Relaxed Density and the Final Gradient}
At each geometry, the fragment solutions are combined into a single global density pair $(\gamma_1,\lambda_2)$ by the amplitude-response assembly described above. Both the reported EWF energy and the nuclear gradient are then evaluated from this same pair, so the energy driving the optimizer's convergence check and the gradient driving its displacement step are self-consistent by construction.

Contracting the assembled density with the electron-repulsion integrals requires care: naively reconstructing the two-particle density as $\gamma_1\otimes\gamma_1 - \tfrac12(\cdots)_{\mathrm{exch}}$ plus $\lambda_2$ introduces a spurious self-interaction term $G[\Delta\gamma_1]$ (the Coulomb and exchange energy of the correlation-only density $\Delta\gamma_1=\gamma_1-\gamma_1^{\mathrm{HF}}$ with itself) that has no counterpart in the EWF energy functional, since that functional couples $\Delta\gamma_1$ only linearly to the (frozen) Fock matrix, never quadratically to itself. We instead build an EWF-consistent effective two-particle density that retains the Hartree--Fock non-cumulant term and both linear cross-terms between $\Delta\gamma_1$ and $\gamma_1^{\mathrm{HF}}$, but omits this spurious term, so that its contraction with the one- and two-electron integrals reproduces $E[\gamma_1,\lambda_2]$ exactly, as an identity valid at any geometry given the same fixed $(\gamma_1,\lambda_2)$. 

The orbital-Lagrangian and CPHF \cite{pople1979derivative} terms are built from this same corrected density; because the corrected contraction and $E[\gamma_1,\lambda_2]$ are identical as functions of geometry, differentiating the corrected contraction through the standard frozen-density CPHF machinery yields the correct orbital-response contribution to $dE/dx$, whereas the naive density would instead differentiate a functional containing the extraneous self-interaction term. With the full molecular-orbital space treated as a single active space, the fixed-denominator (core--active, active--virtual) blocks of the Z-vector vanish identically and only the occupied--virtual CPHF block contributes, so this reduces to the standard closed-shell orbital-response construction applied to the EWF-consistent density. The resulting total gradient combines, in a single contraction, the frozen-density term of Eq.~\ref{eq:split} (including this CPHF orbital response) with the amplitude-response contribution captured through the $\Lambda$-relaxed density\cite{cc_in_quant_chem} described above. We assessed the performance of the $\Lambda$-relaxed density by comparing geometry optimizations of allene carried out with several approaches to assembling the global electron density. These results are presented in the Supporting Information and summarized in Table S1.

\subsection{Geometry Optimization}

\subsubsection{ASE--Sella Interface}
Geometry optimization is performed with Sella,\cite{sella_opt} which subclasses ASE's \texttt{Optimizer}/\texttt{Dynamics} interface\cite{larsen2017atomic}:  the outer convergence loop and force queries are handled by ASE's own \texttt{run()} function, while Sella supplies its own \texttt{step()} function. Sella automatically constructs a basis of redundant internal coordinates from the molecular connectivity and uses iterative Hessian diagonalization to build a partially exact Hessian matrix without requiring the full second-derivative matrix; this Hessian then guides a constrained partitioned rational function approach that steers each step toward the local energy minimum. All optimizations in this work use Sella's default order-0 (minimization) mode with internal coordinates enabled.

As illustrated in Figure~\ref{fig:workflow}, at each optimization step, ASE requests the energy and gradient through a custom calculator. The molecular geometry is passed to the EWF driver, which evaluates the energy and analytical gradient before returning both quantities to ASE. The same calculator, and hence the same optimization loop, is used for the fragmented EWF calculations and for the unfragmented reference calculations reported below: the driver dispatches internally on the calculation mode, so the fragmented and unfragmented geometries are optimized under identical convergence criteria and optimizer settings.

\subsubsection{Optimization Procedure}
The production geometry optimizations were converged when the maximum atomic force fell below $f_{\max} = 0.1$~eV/\text{\AA}, which is the default value used in minimization with Sella\cite{sella_opt}. The benchmarking of global electron density assembly routes, described in Supporting Information, used more robust $f_{\max} = 0.01$~eV/\text{\AA}. Hartree--Fock calculations and clusters solved exactly with FCI were executed on CPUs utilizing modules of the PySCF software package.\cite{sun2018pyscf,sun2020recent} For the large fragments (above 12 MOs), the external SCI or SQD solver, both using the same GPU-accelerated Selected-Basis Diagonalization (SBD) eigensolver, ran on NVIDIA A100 GPUs with libraries configurations as described in the Configuration Interaction Solvers section.

Within each step, fragment-level communication is file based: the driver dispatches one job per fragment for cluster construction and one for the correlated solve, each job reading and writing its own HDF5 file (\texttt{cluster\_<i>.h5}, \texttt{rdm\_<i>.h5}, where \texttt{i} is the index of EWF cluster) so fragments can be solved as independent Slurm-parallel, processes. The assembled global density, energy, and analytic gradient are then computed from these files and handed directly back to the calculator. The per-step geometry and configuration are written to support restarting an interrupted optimization.

%=================================================================
\section{Results and Discussion}

We begin our analysis of EWF geometry optimization performance by examining how closely the geometries optimized with EWF-(FCI,SCI) and unfragmented SCI agree with one another. This comparison allows us to isolate the effect of EWF fragmentation on the quality of the resulting geometries. To quantify this effect, we employ three metrics: the root-mean-square deviation, the largest single-atom displacement, and the number of geometry optimization steps required to reach the equilibrium geometry in each method. 

We use the first 10 molecules from Baker's set, the largest of which, benzene,\cite{Baker_test_set} has 36 MOs in the STO-3G basis set. This choice allows us to perform the unfragmented SCI simulations at a reasonable computational cost, whereas unfragmented SCI simulations of the larger molecules in Baker's set are more computationally challenging and potentially prohibitive. We summarize the results of this analysis in Table~\ref{tab:geom-comparison}.

\begin{table*}
  \centering
  \small
  \caption{Comparison of the optimized geometries obtained from EWF-(FCI,SCI) calculations against the unfragmented SCI reference calculations, for each molecule. Here \textbf{N atoms} is number of atoms, \textbf{RMSD} is root-mean-square deviation between the EWF-(FCI,SCI) and unfragmented SCI calculations, \textbf{Max $\Delta$} is largest single-atom displacement, \textbf{Max EWF MOs} is number of molecular orbitals in the largest EWF cluster, \textbf{N SCI solver} is the number of fragments treated with the SCI solver, \textbf{Full MOs} is the total number of MOs in the unfragmented molecule, and \textbf{EWF steps} and \textbf{Ref. Steps} are number of geometry-optimization cycles in the EWF-(FCI,SCI) and unfragmented SCI runs, respectively.}
  \label{tab:geom-comparison}
  \begin{tabular}{l S[table-format=2.0] S[table-format=2.3] S[table-format=2.3] S[table-format=3.0] S[table-format=3.0] S[table-format=3.0] S[table-format=3.0] S[table-format=3.0]}
    \toprule
    {Molecule} & {N atoms} & {RMSD} & {Max $\Delta$} & {Max EWF} & {N SCI} & {Full} & {EWF} & {Ref.} \\
 & & {(\si{\angstrom})} & {(\si{\angstrom})} & {MOs} & {solver} & {MOs} & {steps} & {Steps} \\
    \midrule
    acetone & 10 & 0.012 & 0.018 & 23 & 4 & 26 & 4 & 4 \\
    acetylene & 4 & 0.016 & 0.021 & 12 & 0 & 12 & 4 & 4 \\
    allene & 7 & 0.033 & 0.040 & 18 & 3 & 19 & 4 & 4 \\
    ammonia & 4 & 0.016 & 0.027 & 8 & 0 & 8 & 5 & 5 \\
    benzene & 12 & 0.017 & 0.021 & 25 & 6 & 36 & 3 & 3 \\
    ethane & 8 & 0.017 & 0.019 & 15 & 2 & 16 & 4 & 4 \\
    ethanol & 9 & 0.020 & 0.025 & 19 & 3 & 21 & 4 & 4 \\
    methylamine & 7 & 0.015 & 0.019 & 14 & 2 & 15 & 2 & 3 \\
    water & 3 & 0.012 & 0.015 & 7 & 0 & 7 & 4 & 4 \\
    \bottomrule
  \end{tabular}
\end{table*}

Table~\ref{tab:geom-comparison} shows that the largest discrepancy between the EWF-(FCI,SCI) and unfragmented SCI optimized geometries is observed in allene, where the RMSD and maximum deviation are 0.033 and 0.040~\si{\angstrom}, respectively. Nonetheless, the overall agreement between the two methods is exceptional across the tested systems, with the majority of molecules exhibiting an RMSD below 0.020~\si{\angstrom} and a maximum deviation below 0.027~\si{\angstrom}. The two methods also agree closely in the number of geometry optimization steps required to reach the equilibrium geometry. This number is identical for all systems except methylamine, where EWF-(FCI,SCI) requires one fewer step to converge than the unfragmented SCI calculation. 

We note that for very small molecules such as water, acetylene, and ammonia, each fragment covers the active space of the entire molecule, and in such instances the benefit of fragmentation diminishes at the chosen bath threshold. Nonetheless, these small cases still serve as useful benchmark points, as they allow us to test the accuracy of the global wave function, electron density, and gradient construction, while adding to the chemical diversity of the tested systems. For these molecules, the small system size keeps the overall MO count below 13 orbitals, which in turn leads to the use of FCI solvers in every fragment—effectively reducing the EWF-(FCI,SCI) scheme to EWF-FCI in these three cases.

\begin{figure*}
  \centering
  \includegraphics[width=\textwidth]{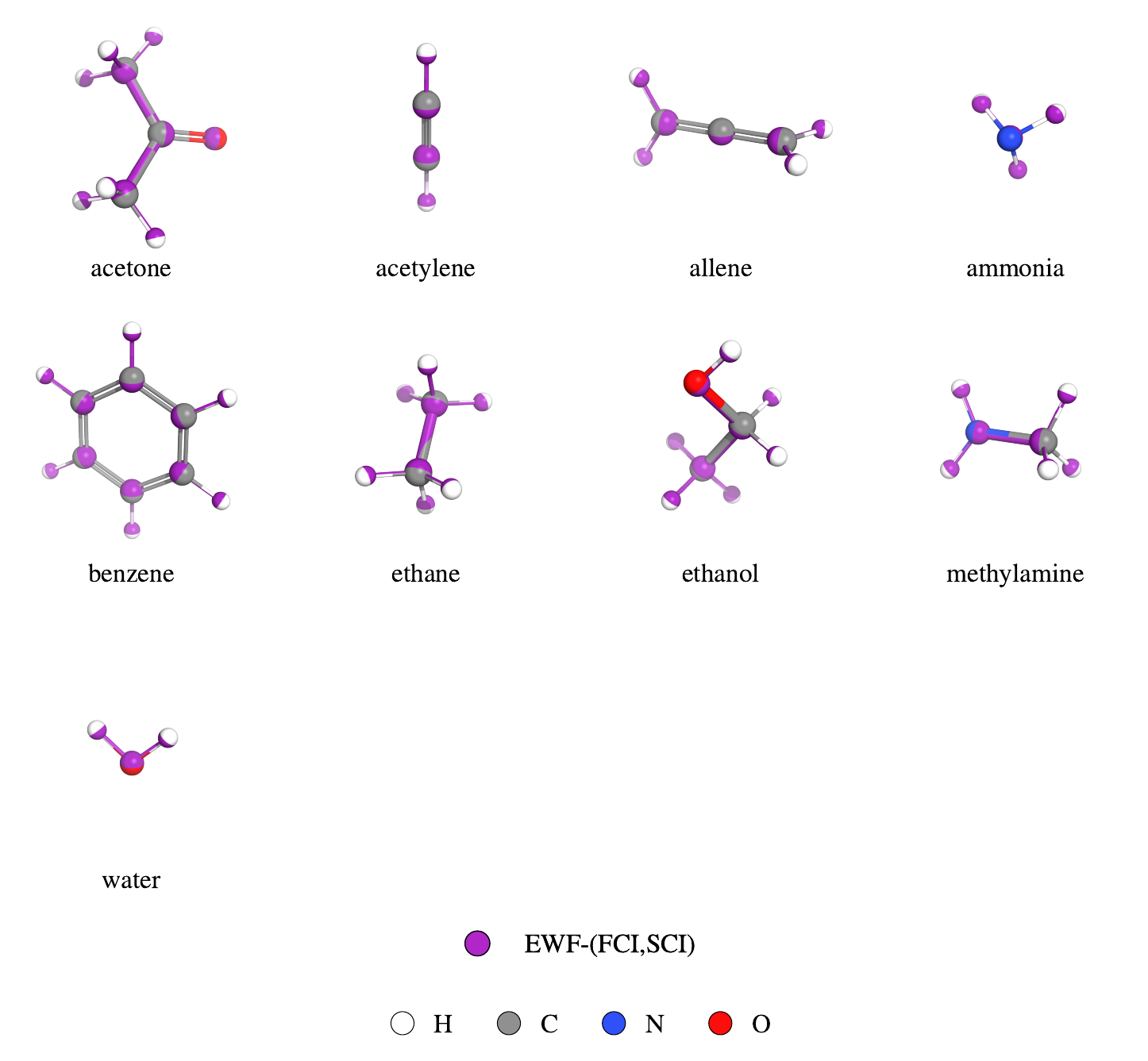}
  \caption{Overlay of the optimized geometries for each molecule. The unfragmented SCI reference is shown in CPK element colors and the EWF-(FCI,SCI) structure in a single highlight color (magenta).}
  \label{fig:overlay}
\end{figure*}

To visually demonstrate how closely EWF-(FCI,SCI) and unfragmented SCI optimized geometries resemble each other we overlay the optimized geometries for all molecules and show them in Figure~\ref{fig:overlay}. As can be seen from Figure~\ref{fig:overlay} the deviations between the geometries produced with EWF-(FCI,SCI) and unfragmented SCI simulations are visually indistinguishable. Such close agreement between the EWF and unfragmented results indicates that fragmentation has little effect on the overall accuracy of the predicted geometries, within the selected set of tested molecules, the chosen convergence threshold of the Sella optimizer, and the limits of the STO-3G basis set.

Next, to demonstrate that the quantum-centric SQD solver can perform as accurately as the classical SCI solver in EWF geometry optimizations, we compare the results of the geometry optimization with EWF-(FCI,SCI) and EWF-(FCI,SQD) methods. We utilize the same three metrics that were used for such comparison in Table 1. We use the same set of test molecules, excluding ammonia, water, and acetylene, and adding two larger molecules, menthone and benzidine, which correspond to the largest and second-largest systems in Baker's test set.\cite{Baker_test_set} Including these two molecules allows us to show that EWF-(FCI,SQD) geometry optimization can be scaled to more practical molecules of importance to biochemical studies.

\begin{table*}
  \centering
  \small
  \caption{Comparison of the optimized geometries obtained from EWF-(FCI,SQD) calculations against the EWF-(FCI,SCI) reference calculations, for each molecule. Here \textbf{N atoms} is number of atoms, \textbf{RMSD} is root-mean-square deviation between the EWF-(FCI,SQD) and EWF-(FCI,SCI) calculations, \textbf{Max $\Delta$} is largest single-atom displacement, \textbf{Max EWF MOs} is number of molecular orbitals in the largest EWF cluster, \textbf{N SQD solver} is the number of fragments treated with the SQD solver, \textbf{Full MOs} is the total number of MOs in the molecule, and \textbf{EWF-(FCI,SQD) steps} and \textbf{EWF-(FCI,SCI) steps} are number of geometry-optimization cycles in the EWF-(FCI,SQD) and EWF-(FCI,SCI) runs, respectively.}
  \label{tab:geom-comparison_q}
  \begin{tabular}{l S[table-format=2.0] S[table-format=2.3] S[table-format=2.3] S[table-format=3.0] S[table-format=3.0] S[table-format=3.0] S[table-format=3.0] S[table-format=3.0]}
    \toprule
     & & & & {Max} & & & {EWF-} & {EWF-} \\
 & & {RMSD} & {Max $\Delta$} & {EWF} & {N SQD} & {Full} & {(FCI,SQD)} & {(FCI,SCI)} \\
{Molecule} & {N atoms} & {(\si{\angstrom})} & {(\si{\angstrom})} & {MOs} & {solver} & {MOs} & {steps} & {steps} \\
    \midrule
    acetone & 10 & 0.000 & 0.001 & 23 & 4 & 26 & 4 & 4 \\
    allene & 7 & 0.014 & 0.018 & 17 & 3 & 19 & 4 & 4 \\
    benzene & 12 & 0.001 & 0.002 & 25 & 6 & 36 & 3 & 3 \\
    benzidine & 26 & 0.007 & 0.010 & 33 & 14 & 82 & 6 & 5 \\
    ethane & 8 & 0.000 & 0.000 & 15 & 2 & 16 & 4 & 4 \\
    ethanol & 9 & 0.000 & 0.000 & 19 & 3 & 21 & 4 & 4 \\
    hydroxysulfane & 4 & 0.000 & 0.000 & 15 & 1 & 16 & 7 & 7 \\
    menthone & 29 & 0.003 & 0.010 & 31 & 11 & 73 & 6 & 6 \\
    methylamine & 7 & 0.000 & 0.000 & 14 & 2 & 15 & 2 & 2 \\
    \bottomrule
  \end{tabular}
\end{table*}

As can be seen from Table~\ref{tab:geom-comparison_q}, the largest discrepancy between the EWF-(FCI,SQD) and EWF-(FCI,SCI) calculations is again observed in allene, where the RMSD and maximum deviation are 0.014 and 0.018~\si{\angstrom}, respectively. In general, the effect of the SQD solver on the overall accuracy of geometry prediction is substantially smaller than the effect of EWF fragmentation, indicating that the SQD solver reproduces the results of the SCI solver remarkably closely. The EWF-(FCI,SQD) and EWF-(FCI,SCI) calculations also agree closely in the number of geometry optimization steps required to reach the equilibrium geometry. This number is identical for all systems except benzidine, where EWF-(FCI,SQD) requires one additional step to converge compared to the EWF-(FCI,SCI) calculation.

\begin{figure*}
  \centering
  \includegraphics[width=\textwidth]{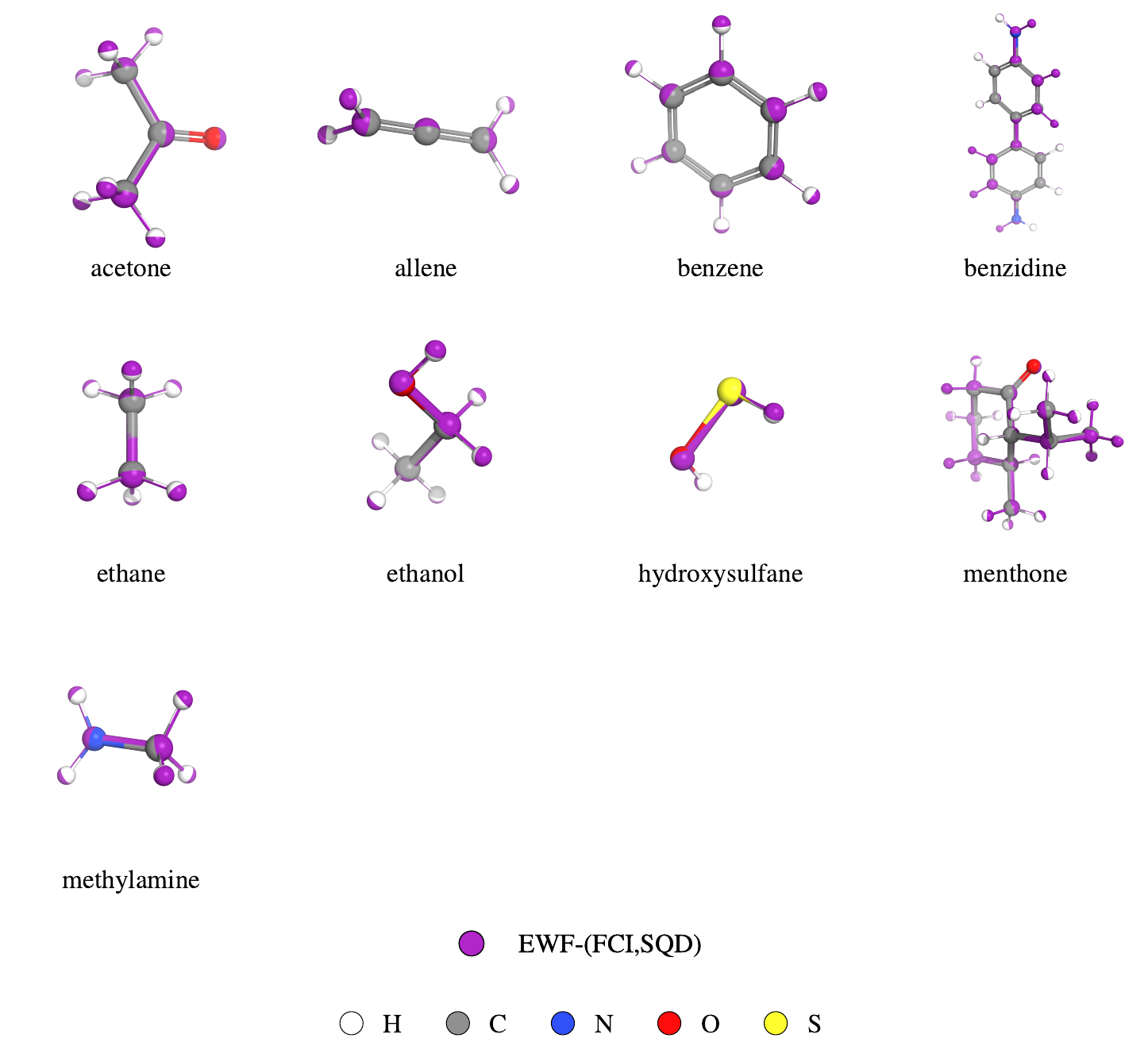}
  \caption{Overlay of the optimized geometries for each molecule. The EWF-(FCI,SCI) reference is shown in CPK element colors and the EWF-(FCI,SQD) structure in a single highlight color (magenta).}
  \label{fig:overlay_q}
\end{figure*}

To visually demonstrate how closely EWF-(FCI,SQD) and EWF-(FCI,SCI) optimized geometries resemble each other we overlay the optimized geometries for 9 molecules represented in Table 2 and show them in Figure~\ref{fig:overlay_q}. As can be seen from Figure~\ref{fig:overlay_q} the deviations between the geometries produced with EWF-(FCI,SQD) and EWF-(FCI,SCI) simulations are visually indistinguishable just like in the comparison performed in Figure~\ref{fig:overlay}. Such close agreement between the EWF-(FCI,SQD) and EWF-(FCI,SCI) results indicates that the choice of SQD solver has little effect on the overall accuracy of the predicted geometries, within the selected set of tested molecules, the chosen convergence threshold of the Sella optimizer, and the limits of the STO-3G basis set. This shows that the SQD solver can be safely used for geometry optimization within the EWF method. While the present study demonstrates the feasibility of using an SQD solver for EWF geometry optimization on real quantum hardware, a demonstration of quantum advantage is beyond its scope. This remains a highly challenging task in the field of quantum computing,~\cite{lee2023evaluating} and one that we intend to address in future studies.

Finally, Table~\ref{tab:geom-comparison_q} characterizes the complexity of the LUCJ quantum circuits used in the SQD simulations within the EWF-(FCI,SQD) geometry optimizations. For each molecule, we report the complexity of the LUCJ circuits for the largest and smallest SQD instances among the EWF clusters, in terms of the number of qubits, the two-qubit-gate circuit depth, and the number of CNOT gates.

\begin{table*}
  \centering
  \small
  \caption{LUCJ quantum-circuit sizes for the smallest and largest EWF clusters treated with the SQD solver, per molecule.  \textbf{SQD Max MOs} / \textbf{SQD Min MOs} are the SQD-treated clusters with the most / fewest molecular orbitals; for each, \textbf{MOs} is the cluster's molecular-orbital (active-space) count, \textbf{Qubits} is the LUCJ ansatz qubit count, \textbf{2Q depth} is the two-qubit-gate circuit depth, and \textbf{CNOTs} is the native two-qubit (CNOT-equivalent) gate count of the transpiled circuit.}
  \label{tab:sqd-circuit-sizes}
  \begin{tabular}{l S[table-format=3.0] S[table-format=3.0] S[table-format=5.0] S[table-format=6.0] S[table-format=3.0] S[table-format=3.0] S[table-format=5.0] S[table-format=6.0]}
    \toprule
    {Molecule} & \multicolumn{4}{c}{SQD Max MOs} & \multicolumn{4}{c}{SQD Min MOs} \\
    \cmidrule(lr){2-5}\cmidrule(lr){6-9}
     & {MOs} & {Qubits} & {2Q depth} & {CNOTs} & {MOs} & {Qubits} & {2Q depth} & {CNOTs} \\
    \midrule
    acetone & 23 & 50 & 108 & 2150 & 16 & 36 & 80 & 1058 \\
    allene & 17 & 38 & 108 & 1232 & 16 & 36 & 80 & 1058 \\
    benzene & 25 & 54 & 116 & 2534 & 25 & 54 & 116 & 2534 \\
    benzidine & 33 & 70 & 148 & 4390 & 17 & 38 & 108 & 1232 \\
    ethane & 15 & 34 & 76 & 934 & 15 & 34 & 76 & 934 \\
    ethanol & 19 & 42 & 118 & 1520 & 13 & 30 & 68 & 710 \\
    hydroxysulfane & 15 & 34 & 76 & 934 & 15 & 34 & 76 & 934 \\
    menthone & 30 & 64 & 136 & 3634 & 17 & 38 & 108 & 1232 \\
    methylamine & 14 & 32 & 72 & 818 & 14 & 32 & 72 & 818 \\
    \bottomrule
  \end{tabular}
\end{table*}

As can be seen from Table~\ref{tab:geom-comparison_q}, all of the SQD simulations use more than 30 qubits, which means that even the smallest SQD simulations in this work exceed the largest VQE simulations demonstrated in the previous DMET-VQE study. The two-qubit-gate circuit depth of the LUCJ circuits ranges from 72 to 148, corresponding to the smallest and largest circuits used in this study—the EWF clusters of methylamine and benzidine, respectively. This range lies within the depths for which quantum circuits are known to maintain good fidelity on IBM quantum hardware where prior benchmarks on IBM processors have demonstrated successful execution of LUCJ circuits with even larger depths and longer durations while remaining safely below the median qubit coherence times ($T_1$ and $T_2$) \cite{merz2026crossing12000atombarrierheterogeneous}. With a shallower depth and thus a shorter execution time, our circuits are well-positioned to maintain high fidelity.

Our recent work has demonstrated that the EWF framework, integrated with advanced subspace solvers, successfully scales to protein systems for electronic energy calculations\cite{trp_cage, merz2026crossing12000atombarrierheterogeneous}. 
A key strength of this underlying architecture is that both the embedding framework and the quantum eigensolvers are systematically improvable.
The robust performance demonstrated in this work opens a clear avenue to deploy EWF-based geometry optimizations across broader for treating larger biochemical systems and exploring more intricate molecular interactions.

%=================================================================
\section{Conclusions}

In this work, we demonstrate the feasibility of EWF-based geometry optimization using twelve molecules from the Baker test set of organic molecules together with the efficient geometry optimizer Sella. Our selection comprises the first ten molecules of the Baker test set and the two largest molecules in the set. We present the implementation of a GPU-enabled SCI solver that uses the software components of PySCF as the main driver for growing the subspace size and SBD as the underlying GPU-enabled eigensolver. To further improve the stability of the calculated energy gradient within EWF, we implement a methodology in which the effective EWF cluster amplitudes are refined by solving the $\Lambda$ equations.

We demonstrate that EWF-based geometry optimization with a mixture of FCI and SCI solvers agrees with unfragmented SCI calculations, with deviations between the EWF and unfragmented geometry optimization results remaining below four picometers. These results indicate that geometry optimization with EWF can be carried out without degrading the overall accuracy through fragmentation, at least within the limits of the selected set of tested molecules, the chosen default convergence threshold of the Sella optimizer, and the STO-3G basis set.

We show that quantum-centric SQD can be deployed as one of the solvers in EWF-based geometry optimizations, scaling these simulations to EWF clusters with up to 33 MOs and using up to 70 qubits. The corresponding EWF-(FCI,SQD) and EWF-(FCI,SCI) geometry optimization results are in excellent agreement, with deviations of only two picometers, which shows that SQD can perform as efficiently as the SCI solver, within the same limits stated above. These results indicate that EWF-(FCI,SQD) emerges as a promising quantum-centric method for exploring potential energy surfaces on real quantum hardware.

Our future studies will address the extension of EWF-(FCI,SQD) geometry optimizations to larger molecules, potentially reaching protein systems. We aim to employ more practical, larger basis sets, and we intend to determine whether further improvements to the EWF methodology can enable the convergence of EWF-(FCI,SQD) simulations in the Sella optimizer within a tighter threshold. Once convergence within such a tighter threshold is achieved, we plan to tackle the geometry optimization of transition states. Finally, the ability of EWF-(FCI,SQD) to produce an electronic gradient that leads to stable convergence signals the strong potential of this method for molecular dynamics simulations, which will constitute another direction of our future work.

%=================================================================

\begin{acknowledgement}
The authors gratefully acknowledge financial support from the National Science Foundation (NSF) through CSSI Frameworks Grant OAC-2209717 and from the National Institutes of Health (Grant Numbers GM130641). The authors are grateful to the high-performance computer center at Cleveland Clinic (Cleveland Clinic Research HPC). We thank George H. Booth for helpful discussion on the EWF method which was essential for identifying optimal approach to deploying geometry optimization EWF. We are grateful to Andreas Goetz for helpful discussion on geometry optimization and gradient stability. We thank Robert Walkup for his help with software setup for GPU-enabled SBD eigensolver.
\end{acknowledgement}

\section{LLM Statement}
Code implementation and debugging was assisted by Claude Opus, Sonnet, and Haiku models under continuous correction, discretion, and review by the authors of this work. Claude Sonnet Opus was used to refine the grammar, punctuation, and sentence structure within the manuscript.

\section{Competing Interest}
The authors declare no competing interest.

\newpage{}
\section*{Supplementary Note 1: Effect of Density Assembly Approach on Geometry Optimization Convergence}

In the embedded wave function (EWF) approach used here, each fragment is solved independently and the resulting per-fragment solutions must subsequently be combined into a single global one- and two-particle density before an analytic nuclear gradient can be evaluated. This combination is not unique. Several prescriptions exist for partitioning the correlation among fragments and for accumulating the per-fragment quantities into the global object, and although all of them are constructed to avoid double counting, they differ in which part of each cluster solution is retained and in how faithfully the assembled density reproduces the derivative of the assembled energy.\cite{Booth2023,nusspickel2022systematic} Because the geometry optimizer consumes both the energy and its gradient, the choice of assembly route affects not only the quality of the optimized structure but also the number of steps required to reach it. This section documents the comparison that informed the choice adopted in the main text. Five routes are compared as implemented in MIQuLab repository~\cite{ewf-geom-opt-repo}.

\textbf{Democratic partitioning} is the most direct prescription. The two-particle cumulant of each cluster is divided among the orbitals that participate in it, so that every correlation contribution is assigned to exactly one fragment, and the resulting pieces are summed directly into the global cumulant. No intermediate wave function is constructed, and the per-fragment density matrices enter the global object essentially unchanged.

\textbf{CI coefficients} takes a global wave function view instead. Each fragment's correlated solution is re-expressed in terms of its single and double excitation coefficients, those coefficients are projected onto the fragment to which they belong, converted into cluster amplitudes, and finally accumulated into one global wave function from which the density is constructed. The appeal of this route is conceptual economy: a single global wave function, rather than a sum of cluster densities, underlies the assembled density. Its limitation is that each cluster solution is truncated to its singles-and-doubles content before assembly.

\textbf{Projected} $\Lambda$ builds the density of each cluster separately, projects each onto its own fragment, and sums the projected densities. This is the default prescription for coupled-cluster-type cluster solvers in the underlying framework.

\textbf{Effective amplitudes} avoids the singles-and-doubles truncation of the CI-coefficient route. Rather than reading excitation coefficients from the cluster wave function, effective single and double amplitudes are extracted from the exact per-fragment density matrices. Because those density matrices are computed from the full selected-CI solution, the renormalizing effect of higher excitations survives into the extracted amplitudes, even though the amplitudes themselves carry only single and double labels. The global density is then assembled from these effective amplitudes using the same coupled-cluster machinery as the CI-coefficient route.

\textbf{Effective amplitudes} $+\ \Lambda$ adds one further step. In the routes above, the Lagrange multipliers required by coupled-cluster density theory are approximated by the amplitudes themselves. Here the corresponding response equations are instead solved explicitly for the assembled global amplitudes, so that the assembled density is the proper response density rather than an approximation to it. This distinction is specific to gradients: solving the response equations makes the assembled energy and its derivative mutually consistent, which is precisely the property a geometry optimizer relies upon when it builds curvature information from successive gradients.

To test these density assembly routes we chose allene as the test system. It is small enough that the full optimization can be repeated once per assembly route alongside an unfragmented reference in the same basis. All calculations reported in this Supplementary Note used the EWF-(FCI,SCI)
method with the STO-3G basis set, with the clusters partitioned between the FCI
and SCI solvers exactly as described in the main text. The SCI
determinant-selection threshold was likewise unchanged, at
$\varepsilon_{\mathrm{SCI}} = 1.0\times10^{-3}$. Geometries were optimized in
internal coordinates subject to a maximum-force convergence threshold of
0.01~eV/\text{\AA},
with the optimizer limited to 25 steps. This threshold is an order of magnitude
tighter than the 0.1~eV/\text{\AA} used for the production calculations
reported in the main text. The reference is an unfragmented selected-CI optimization of the same molecule performed under identical settings, which converged in four cycles.

For each route, the final geometry of the optimization trajectory was aligned to the reference geometry and compared through the root-mean-square deviation and the largest single-atom displacement. The number of optimization cycles is reported alongside, since a route that reaches an accurate structure only after many steps is of limited practical value for larger systems.

\begin{table*}
  \centering
  \small
  \caption{Effect of the EWF density-assembly route on the optimized geometry of allene. \textbf{RMSD} is the root-mean-square deviation and \textbf{Max $\Delta$} the largest single-atom displacement of the final geometry from the unfragmented SCI reference; \textbf{Steps} is the number of geometry-optimization cycles. The assembly routes are: \textbf{Democratic} --- four-index democratic partitioning of the per-fragment two-particle cumulant;  \textbf{CI coefficients} --- per-fragment CISD coefficients are projected onto the fragment, converted to cluster amplitudes, and tiled into a global wave function;  \textbf{Projected $\Lambda$} --- sum of single-cluster projected cumulants;  \textbf{Effective amplitudes} --- effective $(T_1, T_2)$ amplitudes extracted from the exact per-fragment cumulant;  \textbf{Effective amplitudes $+\ \Lambda$} --- as for the effective amplitudes, with the $\Lambda$ (Z-vector) equations solved for the assembled global amplitudes.}
  \label{tab:si-assembly-modes-allene}
  \begin{tabular}{l S[table-format=1.3] S[table-format=1.3] S[table-format=2.0]}
    \toprule
    {Assembly mode} & {RMSD} & {Max $\Delta$} & {Steps} \\
     & {(\si{\angstrom})} & {(\si{\angstrom})} & \\
    \midrule
    Democratic & 0.004 & 0.004 & 6 \\
    CI coefficients & 0.006 & 0.009 & 25 \\
    Projected $\Lambda$ & 0.038 & 0.045 & 5 \\
    Effective amplitudes & 0.026 & 0.032 & 4 \\
    Effective amplitudes $+\ \Lambda$ & 0.020 & 0.024 & 4 \\
    \bottomrule
  \end{tabular}
\end{table*}

The results in Table~\ref{tab:si-assembly-modes-allene} separate two properties that are easily conflated: how close a route brings the optimized structure to the unfragmented reference, and how readily the optimizer reaches a stationary point at all. No single route is best on both counts for this system.

The two effective-amplitude routes converge fastest. Both reach the force threshold in four cycles, exactly matching the unfragmented reference, so for this system the fragmentation imposes no penalty. Solving the response equations improves the structure at no additional cost in steps: the root-mean-square deviation falls from \SI{0.026}{\angstrom} to \SI{0.020}{\angstrom} and the largest single-atom displacement from \SI{0.032}{\angstrom} to \SI{0.024}{\angstrom}. This is the behavior expected from a properly relaxed response density, whose gradient is consistent with the energy it is derived from.

The CI-coefficient route did not reach the force threshold within the 25-step limit. It is worth stating explicitly that this does not indicate a defect in the route. Its deviation from the reference at the point of termination is small --- \SI{0.006}{\angstrom} --- so the optimization is in the correct region of the potential energy surface and is not diverging; what it cannot do is reduce the residual forces below the requested threshold. A plausible explanation is the truncation inherent to this route: each cluster solution is reduced to its singles-and-doubles content before assembly, and in a minimal basis that content is a comparatively poor proxy for the full cluster solution, leaving a residual component in the assembled gradient that the optimizer cannot remove. This limitation is specific to the regime probed here. In a larger basis, under a different fragmentation, or for systems where the global wave function picture is better justified, this route may well be the preferable choice, and nothing in the present test argues otherwise. The deviations quoted for this row necessarily refer to an unconverged geometry and should be interpreted with that in mind.
Democratic partitioning gives the closest agreement with the unfragmented reference of any route tested, with both the root-mean-square deviation and the largest displacement at \SI{0.004}{\angstrom}. Its convergence is nevertheless slower, requiring six cycles against four for the effective-amplitude routes. Accuracy of the final structure and efficiency of the path toward it are therefore not the same axis, and a route selected on either criterion alone would be chosen for the wrong reason.

Finally, the present comparison is expected to understate the advantage of the effective-amplitude routes. Their distinguishing feature is that the amplitudes are sourced from the exact cluster density matrices rather than from a singles-and-doubles wave function, and the benefit of doing so grows as the cluster solution departs from a singles-and-doubles description --- that is, for stretched bonds, near-degeneracies, and larger clusters whose selected-CI subspace approaches the full configuration interaction limit. Allene near its equilibrium geometry in a minimal basis is a weakly correlated case, in which the effective amplitudes lie close to their coupled-cluster counterparts. A more strongly correlated test system would be expected to widen the difference between these routes and the alternatives.

The objective of the present study is to demonstrate that geometry optimization is feasible when the cluster solutions are obtained by sample-based quantum diagonalization. The comparison reported here served the narrower purpose of selecting a density-assembly route for that demonstration, and the effective-amplitude route with the response equations solved was adopted in the main text as offering the most favorable balance between the accuracy of the optimized structure and reliable convergence of the optimization. 

That choice should not be read as a general ranking of the assembly routes. The comparison above involves a single molecule in a minimal basis at a single convergence threshold, and the relative merits of the routes plainly depend on basis set size, correlation strength, and the fragmentation scheme. A systematic study establishing which route is preferable for which class of system remains necessary, but it addresses a different question from the one posed here and is therefore left to future work. 

%\bibliography{references_SI}

\newpage
\bibliography{references}

\end{document}